\documentclass[graybox]{svmult}

\usepackage{mathptmx}       
\usepackage{helvet}         
\usepackage{courier}        
\usepackage{makeidx}         
\usepackage{graphicx}        
\usepackage{multicol}        
\usepackage{natbib}

\makeindex             

\newcommand{\aj}{AJ}
\newcommand{\apj}{Ap.J}
\newcommand{\apjl}{Ap.J Letters}
\newcommand{\apjs}{Ap.J Supplement}
\newcommand{\aaps}{AAPs}
\newcommand{\araa}{ARA\& A}
\newcommand{\mnras}{MNRAS}
\newcommand{\aap}{A\& A}
\newcommand{\apss}{Ap \& SS}

\newcommand{\re}{r_{\rm e}}
\newcommand{\mue}{\mu_{\rm e}}
\newcommand{\nb}{n_{\rm b}}
\newcommand{\Dn}{D$_{\rm n}$(4000)}
\newcommand{\bn}{b_{\rm n}}

\begin{document}

\title{An Observational Guide to Identifying Pseudobulges and
  Classical Bulges in Disk Galaxies}
\titlerunning{ Identifying Pseudobulges and
  Classical Bulges} 
\author{David B Fisher and Niv Drory}

\institute{David B. Fisher \at Swinburne University of Technology, \email{dfisher@swin.edu.au}
\and Niv Drory\at University of Texas, \email{drory@astro.as.utexas.edu}}

%
%
\maketitle

\abstract{In this review our aim is to summarize the observed
  properties of pseudobulges and classical bulges. We utilize an
  empirical approach to studying the properties of bulges in disk
  galaxies, and restrict our analysis to statistical properties. A
  clear bimodality is observed in a number of
  properties including morphology, structural properties, star
  formation, gas content \& stellar population, and kinematics. We 
  conclude by summarizing those properties that isolate pseudobulges
  from classical bulges. Our intention is to describe a practical,
  easy to use, list of criteria for identifying bulge types. }

\section{Introduction}
\label{sec:1}
This paper reviews those observed properties of bulges that reveal the
bimodal nature of the central structures found in disk galaxies. Our
aim is to collect a set of empirical properties of bulges that can be
used to diagnose bulges into the two subcategories commonly referred
to as {\em pseudobulges} and {\em classical bulges}. Despite a long
history of studying bulges in disk galaxies \citep{hubbleatlas}, and
the knowledge that bulges are very common, being found in upwards of
$\sim$80\% of bright galaxies ($>10^9$~M$_{\odot}$;
\citealp{fisherdrory2011}), only recently have systematic studies of
the bimodal nature of bulges become frequent in the literature.

Kormendy \& Illingworth (1983) have shown that bulges in disk galaxies
separate by internal kinematics: some rotate rapidly like a disk where
others are dominated by random motions
\citep{kormendy_illingworth_1982}. Also, the review by
\citet[][and references therein]{wyse1997}  demonstrates clearly that
bulges are a heterogeneous class of objects. Bulges are shown to vary
significantly in their ages and metallicities, and not all bulges show
properties that are similar to elliptical galaxies.  The observation
that there is more than one type of bulge introduces the possibility
that bulges as a class could be the end result of more than one
mechanism of galaxy evolution.

\begin{figure}
\begin{center}
\includegraphics[width=0.85\textwidth]{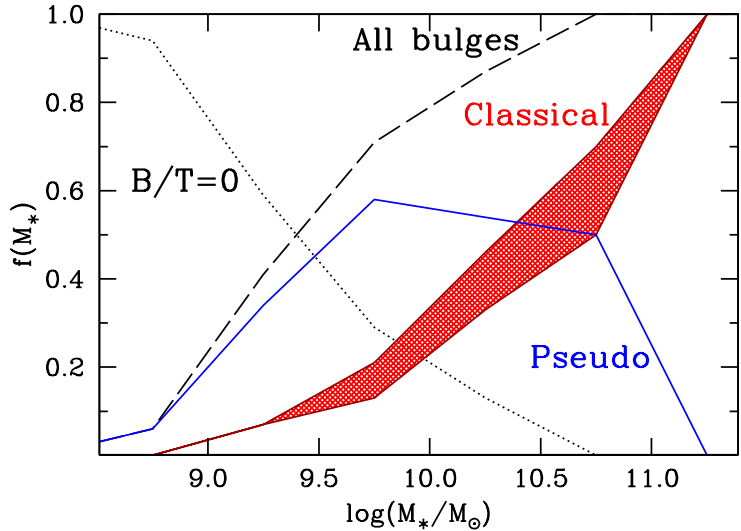}
\end{center}
\caption{The frequency of bulge types correlates with total galaxy
  mass. The four curves indicate the frequency of pseudobulges (blue
  solid line), classical bulges (red filled region), galaxies with no
  bulge (dotted line) and all bulges (dashed line) as a function of
  total galaxy mass. The classical bulges are shown as a shaded region
  because an attempt has been made to account for composite
  pseudobulge-classical systems. The higher value for a given
  mass includes this estimate, the lower value is for galaxies whose
  bulges are pure-classical bulge systems.  There is a clear sequence
  of bulgeless galaxies existing at low mass, pseudobulges in
  intermediate mass galaxies and classical bulges in high mass
  galaxies. \label{fig:freq}}
\end{figure}
In Fig.~\ref{fig:freq} we show a result that illustrates simple
evidence that bulge type is connected to the evolution of
galaxies. The figure shows the frequency of bulge types for the
brightest $\sim$100 galaxies in the local 11~Mpc volume. The type of
bulge a galaxy contains changes systematically as galaxy mass
increases. Similarly, galaxies with blue, young, stellar populations
have been shown to have very different bulges than those of red, old
galaxies \citep{droryfisher2007}. These results suggest that bulge
type is connected to the phenomena that drive galaxy evolution. Being
able to diagnose bulge types in galaxies is therefore both useful to
understand the properties of an individual galaxy, and also to
understand galaxy evolution in general.

At present, we know of three main mechanisms that allow a galaxy to
grow bulge mass (as measured by an increase in the bulge-to-total
luminosity ratio from bulge-disk decompositions). These are merging
processes \citep{hammer2005,aguerri2001}, slow secular evolution
\citep{kk04,athan2005}, and rapid internal evolution due to disk
instabilities during the ``clumpy'' phase
\citep{elmegreen2008,inoue2012tmp}.  It is therefore critical that we
be able to identify the properties of bulges that potentially isolate
features associated with each of these formation channels. Given that
realistically the end result of bulge formation and evolution is
likely a composite object, recognizing ``pure'' examples of each
formation channel (i.e.\ the most extreme cases along a spectrum of
properties) will be necessary to disentangle the physical processes
involved.

In this review we will concentrate on work separating bulges into the
dimorphic classes mentioned above. These two categories have been
given names, the most popular of which seem to be ``pseudobulges'' and
``classical bulges''. In short, pseudobulges are bulges that have
properties that historically we associate with dissipative phenomena
(active star formation, rotating kinematics, young stars;
alternatively, some authors refer to such bulges as ``disky bulges''
\citep{athan2005}). \citet{kk04} give a
thorough review, though now 10 years old, of pseudobulge
properties. That review focuses largely on exemplary cases, while the
review here will focus on statistical results, which can be applied to
large sets of galaxies. ``Classical bulges'', in turn, are those
bulges that exhibit properties resembling elliptical galaxies, such as
smooth distribution of stars, old stellar age, and kinematics
dominated by random motions. The term ``classical'' refers to this being
the widespread preconception about bulges for much of the 20th century
\citep{wyse1997}. Using a terminology that is based on preconceptions
that are no longer widely held seems a bit archaic. Nonetheless, we
accept the concept in language signification (known as {\em Saussurean
  Arbitrariness}), in which historic meaning or sound of a word is not
as important as the meaning we ascribe to it now, and simply adopt the
most popular terms of the present day (``pseudobulges'' and
``classical bulges''). For further reading on bulge properties we
refer the reader to the aforementioned reviews by Kormendy \&
Kennicutt (2004) and Wyse et al.\ (1997), and also the lecture notes
by \citet{gadotti2012} and \citet{kormendy2013}.

\subsection{Definition of a Bulge}

Before discussing the separate kinds of bulges, it is necessary to
define what is meant by the term ``bulge'' when applied to galaxies.
The most commonly used definition of bulges is based on the observed
rise in surface brightness above the disk that is observed at the
center of many intermediate-type galaxies. Disk components of galaxies
are often well described by an exponential decay with increasing
radius of their surface brightness \citep{freemanlaw}.  Many galaxy's
contain a centrally located structure that is brighter than the inward
extrapolation of the disk's exponential surface brightness, and this
component is not associated with a bar. This central structure is
often identified as a ``bulge''.  Bulges of this type are often
identified using bulge-disk decomposition techniques
\citep{kormendy1977decomp}, commonly using the S\'ersic function
\citep{sersic1968} to describe the surface brightness of the
bulge. Defining bulges using surface photometry has the advantage that
it is straightforward, empirically based, and can be applied to large
numbers of galaxies. In principle one can use large data sets like the
Sloan Digital Sky Survey to characterize bulges in $>10^4$ galaxies
\citep{lackner2012}.

Identifying bulges in bulge-disk decomposition is by nature parametric,
but disk galaxies could have non-exponential components in their
centers (similar to bars). Therefore, identifying extra light as a
seperate component may be misleading and physically meaningless.  An
alternate view of this is that in some intermediate-type galaxies, the
bulge-disk decomposition simply reflects an empirical description of
the surface brightness profile of star light.  Another weakness of
this method is that bulge-disk decomposition using the S\'ersic
function (described below) appears deceptively simple, yet the
procedure carries with it a high degree of degeneracy.

Bulges are also identified as a 3-dimensional structure that
``bulges'' from the disk plane in the $z$ direction. These structures
are most easily identified in edge-on galaxies where bulging central
structures are observed in the vast majority of massive galaxies
\citep{kautsch2006}. A significant caveat, however, to studying bulges
in edge-on systems is that dust extinction from the disk significantly
affects the light of the bulge, especially in galaxies with smaller
bulges. Secondly, boxy bulges \citep{bureau1999} which are the result
of bars \citep{athan05} can complicate the interpretation of bulge
thickness. Two edge-on galaxies could have equally thick centers one
with a boxy-bulge the other with a round thick bulge, which would be
missed by blanket thickness cuts.

Kinematics can be used to identify a low-angular momentum and higher
$z$-dispersion structure at the center of a high angular momentum thin
disk.  For example \citet{fabricius2014} show that kinematics of the
intermediate-type galaxy NGC~7217 clearly separates into two
components one with high dispersion (the bulge) and the second with
low dispersion (the disk). These components are consistent with a
photometric bulge-disk decomposition. Ideally such procedure could be
carried out on large numbers of galaxies in forthcoming data releases
of SAMI \citep{croom2012} and MaNGA \citep{Bundy2014}. However, its
not clear that either survey has sufficient spatial or spectral
resolution to apply this technique.

\subsection{Outline}

Kormendy \& Kennicutt's (2004) review and Athanassoula (2005) make a
strong case that multiple types of bulges exist, and that this is
likely reflecting different channels of bulge formation and galaxy
evolution.  In this review, we discuss the identification of bulges of
different types, attempting to provide practical means of classifying
bulges.

\begin{figure}
\begin{center}
\includegraphics[width=0.9\textwidth]{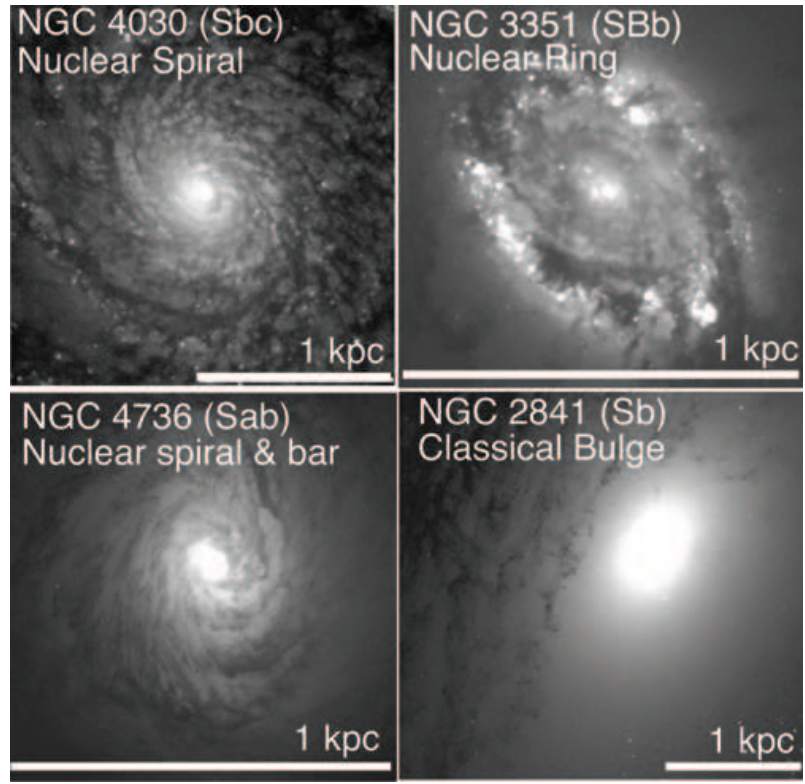}
\end{center}
\caption{ Examples of bulge morhologies are shown using optical images
  from HST. The detectors and filters are NGC~4030: PC F606W;
  NGC~3351: PC F606W; NGC4736: PC F555W; and NGC~2841: ACS/WFC F435W.
  The white line in each panel represents 1~kpc. There is an extreme
  difference in these galaxies between pseudobulge morphologies
  (nuclear ring, spiral and bar) and classical bulges. In cases such
  as this, morphological diagnosis of bulge types is relatively
  straightforward.  \label{fig:morphology}}
\end{figure}

\section{Identifying Pseudobulges with Morphology}

There are multiple lines of reasoning that motivate the morphological
distinction of different bulge types. First, empirically speaking,
results from {\em Hubble Space Telescope} (HST) imaging surveys are
quite clear that there is not one single type of morphology that can
be associated with regions of galaxies dominated by bulge light. This
is in contrast to the description of bulges given in the Carnegie
Atlas of Galaxies \citep{carnegieatlas}, in which bulges are described
as having no evidence of a disk or ``pure E'' morphology. The presence
of spiral structure (see, for example, Fig.~\ref{fig:morphology}) is
in stark contrast to this definition. If the structure exhibiting the
spiral, ring, or bar pattern is dominating the light then the
classifier can be fairly confident that the dynamical state of the
system better reflects that of disk kinematics than that of an
elliptical galaxy. Morphology is therefore a physically motivated
classification. However, we have to remind the reader of the problem
in identifying such a disky structure as a distinct component as
opposed to just being the physical state of the central disk.

From a certain point of view the simplest means of identifying bulges
of different types is morphology. The main requirement is sufficient
spatial resolution to identify small-scale features. Data from HST has
made this a very straightforward process in which high quality
identification of features like spiral structure can be done on nearby
galaxies ($<50$~Mpc). Typically, in this practice the user identifies,
by-eye, features that are associated with disk morphology (such as
spirals, rings, and bars) inside the region where the bulge dominates
the light of the galaxy.  Systematic studies comparing
morphological bulge classification at different wavelengths would be
useful. It stands to reason that broadband photometry at
wavelengths in the middle of the optical spectrum (ie. $V$ to $I$) are
best suited. If the filter is too blue, the light becomes too
sensitive to dust effects. Although, it has been shown that the
morphological features identifying pseudobulges are present in near-IR
images \citep{fisherdrory2010}, these features become difficult to see
at longer wavelengths (eg. $JHK$ bands).

Results from HST reveal that the centers of relatively ``early type''
galaxies (Sa-Sb) frequently contained spiral structure and show little
evidence of a smooth featureless bulge \citep{carollo97}. In
Fig.~\ref{fig:morphology}, top left panel, we show an example of
nuclear spiral morphology. In this example, NGC~4030, the spiral is
face-on and quite easy to identify.  When present, the spiral
structure frequently extends throughout the entire bulge, and reaches
to the very center of the bulge region. In the centers of later type
galaxies, such dusty spiral and non-smooth morphology becomes much
more common than smooth, round bulges \citep{boeker2002}.  In very
nearby galaxies, e.g.\ NGC~5055, the presence of spiral structure that
extends all the way to galaxy centers was recognized as early as 1961
in the Hubble Atlas \citep{hubbleatlas}.  \citet{buta1993}
identify a sample of nuclear spirals which they call pseudorings,
placing first estimates on sizes (typical diamaters of $\sim$1~kpc).
The advent of surveys from HST make it clear that nuclear spirals are
very common in Sa-Sm galaxies \citep{fisherdrory2011}. Fisher \& Drory
(2008) introduce a secondary category of spirals referred to as
nuclear patchy spirals. These are almost exclusively found in later
type (Sc-Sd) galaxies with very small bulges.

A ``nuclear ring'' is a ring of stars and/or intense star formation
found in the central region (radius $<$~1~kpc) of a disk galaxy
\citep{buta1993,buta2007}. Nuclear rings are often relatively easy to
identify, and they are typically very bright due to their large star
formation rates.  Nuclear rings are separate from ``inner rings'' that
are commonly found at the end of bars \citep{rc3}. Nuclear rings occur
in roughly 20\% of spiral galaxies \citep{knapen2005}. Galaxies with
nuclear rings are very likely to be barred
\citep{comeron2010,knapen2005}. In Fig.~\ref{fig:morphology} we show
an example of a prominent nuclear ring in the nearby disk galaxy
NGC~3351. \citet{buta1993} identify galaxies with both nuclear
rings and ``pseudo-rings''. A pseudo-ring is when the ring is not
fully formed, and does not extend 360 degrees around the galaxy
center.  In fact, it may commonly refer to nuclear spirals.

Studies focusing on barred spirals find that secondary (nested) bars
are frequent \citep{erwin2002,erwin2004}. As many as 40\% of S0-Sa
galaxies with bars contain a secondary bar, extending to radii of
0.2-0.8~kpc. Many of the studies on secondary bars focus on early-type
galaxies where there is less dust and the bars are easier to
identify. Secondary bars in later-type galaxies are easily obscured by
dust, and often hard to identify for that reason. Even in unobscurred
galaxies, it is useful to over-plot isophote contours of the galaxy to
identify nuclear bars (as outlined by Erwin \& Sparke 2002, also Erwin
2004). In Fig.~\ref{fig:morphology} (bottom left panel) we show a
galaxy with both a nuclear spiral and a nuclear bar. The bar is
aligned north-to-south in the image. A number of simulations focus on
the formation of galaxies with nested bars
\citep{heller2007a,debattista2007,shen2009}. These simulations
generally find that the nuclear bars are rapidly rotating structures
that form easily within barred disks.

Classical bulges are morphologically identified, in the ideal case, as
having smooth centrally peaking isophotes that do not show any
evidence of disk-like structure such as those described above. In
Fig.~\ref{fig:morphology} we show NGC~2841 as an example. In the image
the smooth classical bulge is seen in the center, and at larger radii
the effects from the disk become apparent. The presence of some
extinction, indicating dust and gas, does not preclude a system from
being a classical bulge; however in classical bulges when defined by
morphology, such dust is not a dominant feature, nor is it embedded in
a spiral pattern.

There are a number of caveats associated with morphological
classification of bulge types.  Using morphology as a means of
identifying physically distinct phenomena is an inherently biased
process by the person doing the identification. Two individuals can
come to different conclusions about what is or is not a spiral
pattern, or just a wisp of dust. Even with HST data, morphological
classification is only possible at very low redshifts
$z<0.05$. Finally, in the absence of Galaxy Zoo type of analysis
\citep[e.g.][]{lintott2011} morphology is not a quantitative science;
this limits both our ability to interpret the meaning and also to
apply such analysis to large samples of objects.

Combining all disk-like structures (nuclear rings, nuclear spirals,
and nuclear bars) into a single category of ``pseudobulges'' makes the
assumption that these objects are linked. The conditions under which
nuclear rings form are likely different than that of a secondary bar,
nonetheless, the unifying concept is that all three are structures that
are associated with disks. Furthermore, there is no significant
differences between the bulge S\'ersic index, bulge-to-total ratio or
half-light radius of bulges with these structures
\citep{fisherdrory2008}. The strongest difference appears to be
between classical bulges and the rest of bulge morphologies.

In spite of the many caveats, morphological identification of bulge
types seems quite useful. Bulges identified as pseudobulges using
morphology are more actively forming stars \citep{fisher2006}, have
more disk-like kinematics \citep{fabricius2012}, and occupy a
different location in structural parameter space
\citep{fisherdrory2010} than classical bulges. These correlations will
be discussed in more detail in subsequent sections, since their
existence does establish that by-eye classification can accurately
mark important distinctions.

\section{Structural Properties of Bulges: S\'ersic index, Scaling
  Relations, and Shape of Bulges}

Structural parameters returned from bulge-disk decompositions can be a
very powerful means to identify pseudobulges. In theory, bulge-disk
decomposition software can be run on very large numbers of
galaxies. If one can robustly identify bulge-types from the properties
in decompositions alone, it is then straightforward to generate strong
constraints on the number of bulges of each type in different
environments. In practice, this procedure is complicated by inherent
degeneracies in the decomposition procedure.

The process of bulge-disk decomposition assumes that the radial
surface brightness profile, $I(r)$, of a galaxy can be described by a
linear combination of a small number of component structures, such
that $I(r) = I_{\rm bulge}(r) + I_{\rm disk}(r) + I_{\rm other}(r)$, where
$I_{\rm bulge}$ and $I_{\rm disk}$ describe the bulge and disk, and
$I_{\rm other}$ describes any other structure in a galaxy.

There are a few systematic sources of uncertainty that should be taken
into account to derive accurate parameters from bulge-disk
decompositions. First, a well-known problem is accounting for galaxy
structures that are neither bulge nor an exponential disk. Most
commonly $I_{\rm other}$ describes light from a bar, but could also refer
to rings, nuclei, or bright star forming spiral arms. Not taking a bar
into account when modeling the light profile leads to systematic
effects, such as overestimating the bulge-to-total ratio ($B/T$) by as
much as a factor of two, and also systematically overestimating the
value of the S\'ersic index \citep{gadotti2008,fisherdrory2008,laurikainen2006}. If
the galaxy has a central point source, either AGN or nucleus, this
must be accounted for as well or else the returned model will have an
artificially large S\'ersic index and $B/T$.

Resolution is a crucial parameter for determining accurate S\'ersic
model parameters of bulge-disk decompositions. If the bulge is of the
size of the resolution element, information on the size (half-light
radius) and shape (S\'ersic index) are completely untrustworthy
\citep{gadotti2008,fisherdrory2008}. \citet{gadotti2008} suggests that
at least 80\% of the half-light radius must be
resolved. \citet{fisherdrory2010} find that in order to determine
accurate S\'ersic indices of galaxies with small B/T, a resolution of
100~pc is preferred.

\subsection{Using S\'ersic Index to Identify Bulge Types}

Typically, bulge-disk galaxies are decomposed using the S\'ersic function
\citep{sersic1968} to describe the bulge. In this model the radial
light profile in units of mag~arcsec$^{-2}$ of a galaxy can be described as 
\begin{equation}
\mu_{\rm bulge}  = \mu(\re) + \bn \left[ \left(\frac{r}{\re}\right)^{1/\nb} -1 \right], 
\end{equation}
where $\nb$ is the S\'ersic index of the bulge; $r$ represents radius,
$\re$ is the radius containing half the light of the bulge, $\mu(\re)$
is the surface brightness at $\re$, and $\bn=2.17 \nb - 0.355$. The
above formula, known as the S\'ersic function \citep{sersic1968}, has
been shown by a number of authors to describe the shape of elliptical
galaxy profiles quite well \citep{caon94}. The case of a S\'ersic
function with $\nb=1$ is equivalent to an exponential, commonly used
to describe disks. The case of $\nb=4$ is equivalent to the
de~Vaucaleurs profile used historically for E-type galaxies. There are
a number of detailed discussions of the S\'ersic function and its
properties; for further reading see \citet{graham2005}.

The S\'ersic index of bulges is now widely used as a means to identify
bulges, based largely on its correlation with other bulge properties
\citep[e.g.][]{fisherdrory2008}.  The first evidence came from early
surveys of bulge-disk decomposition in which it was clear that many
bulges are better fit by double exponential profiles than by a
traditional de~Vaucaleurs profile \citep{andredak94,courteau96}. This
was eventually generalized to show that all bulges are better
described using the S\'ersic funciton \citep{andredak95}, and that
later-type galaxies tend to have lower values of $\nb$. Results using
HST images find that bulges with disk morphology are more likely to
have shallow, more like exponential, surface brightness profiles
\citep{scarlata2004}.

Using $\sim$100 galaxies with HST imaging, \citet{fisherdrory2008}
compare the morphology of bulges to the associated bulge S\'ersic
index from detailed bulge-disk decompositions. They find that there is
a clear bimodal distribution of S\'ersic indices in galaxies.  To
reduce uncertainty in the S\'ersic index, the authors created
composite surface photometry using HST data to measure the surface
brightness profile of the inner 10 arcsec, and a set of deep
wide-field images to measure the surface brightness profile of the
outer parts of the galaxy (a similar procedure is discussed in
\citealp{balcells2003} and \citet{kfcb}). The result is a surface
brightness profile that covers a very large dynamic range in radius,
and is thus able to reduce uncertainty in S\'ersic index, and better
break the degeneracy between $n$ and $\re$ \citep{graham96}. These
decompositions reveal that 90\% of bulges with morphology that
indicates a pseudobulge (as described in the previous section) have
$\nb<2$, and all classical bulges and elliptical galaxies have
$\nb>2$. The authors followed up on this result with a larger sample
of galaxies with near-IR photometry (still combining HST and in this
case Spitzer IRAC~3.6~$\mu$m; \citealp{fisherdrory2010}). The result
is the same, all classical bulges are found to have $\nb>2$ and over
90\% of pseudobulges have $\nb<2$.

\begin{figure}
\begin{center}
\includegraphics[width=0.9\textwidth]{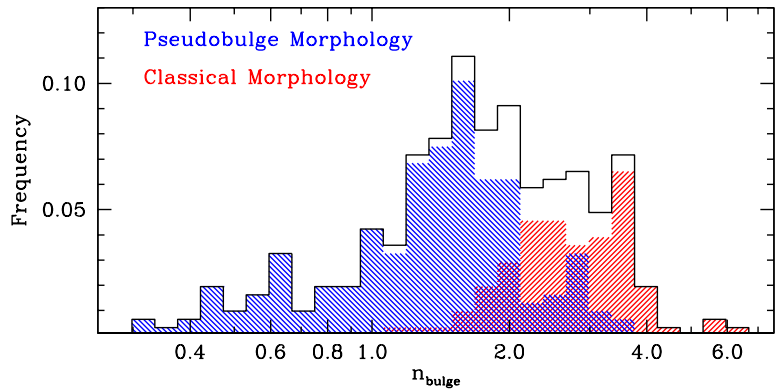}
\end{center}
\caption{The distribution of bulges S\'ersic indices from a sample of
  308 nearby bulge-disk galaxies with both published bulge-disk
  decompositions and avaliable data in the HST archive for bulge
  morphology diagnosis. The distribution of $n_{\rm bulge}$ in galaxies
  with classical bulge morphology shown to be clearly different than
  that of pseudobulges.  \label{fig:sersic}}
\end{figure}

To double check the correlation of bulge S\'ersic index with high
resolution bulge morphology we compile a sample of 308 galaxies that
have both published bulge-disk decomposition and also have data in the
HST archive from which we can determine the morphology of the
bulge. The sources of $\nb$ are
\cite{fisherdrory2008,fisherdrory2010,fisherdrory2011,fabricius2012,fisher2013,laurikainen2010,weinzirl2009}. In
the case of overlapping galaxies we take the result that is based on
the finest spatial resolution, though typically the spread in S\'ersic
index is not large, $\Delta \nb<0.2$. In a few galaxies ($\sim 10$)
the spread in $\nb$ is large, $\Delta \nb>1$. We drop these galaxies
assuming that the S\'ersic index is poorly constrained and not
trustworthy. The total sample combines decompositions from three
independent fitting procedures \citep[described
in][]{fisherdrory2008,weinzirl2009,laurikainen2010}, and contains 106
S0-S0/a, 71 Sa-ab, 62 Sb-bc, 61 Sc-cd, 10 Sd-dm galaxies.

In Fig.~\ref{fig:sersic} we show the distribution of S\'ersic indices
in the combined sample. There is a clear correlation between bulge
type and S\'ersic index. The choice of $\nb=2$ as the dividing line is
not arbitrary, but rather is justified by the coincidence of this
value with the turnover in the two distributions. This is clearly
evident in the figure. The sample contains 102 classical bulges and
87\% of those classical bulges have $\nb>2$, conversely in the sample
we identify 205 galaxies as having pseudobulges and 86\% of these have
$\nb<2$. If we consider only those galaxies with Hubble type Sa and
later, the frequency of classical bulges with $\nb<2$ drops to 7\%,
and the frequency of pseudobulges with larger S\'ersic index ($\nb>2$)
becomes only slightly lower, 11\%.  We note that although not
completely devoid of gas, S0 galaxies have significantly less dust and
gas \citep{young2011}, therefore identifying features such as nuclear
spirals is much more difficult in these galaxies.

\begin{figure}
\includegraphics[width=0.85\textwidth]{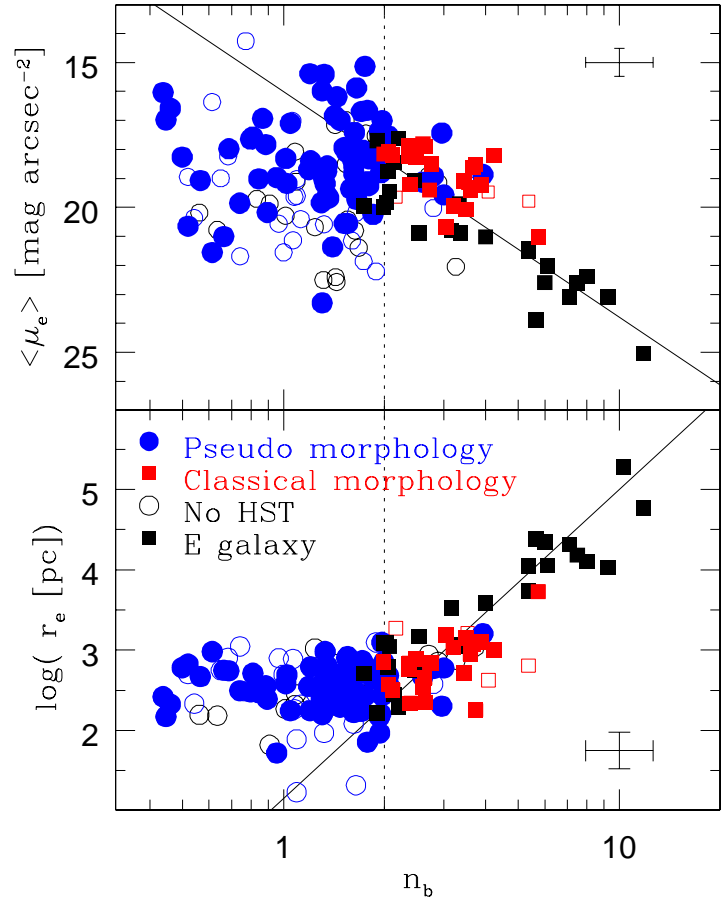}
\caption{The above figure, adapted from \citet{fisherdrory2010}, shows
  the correlation of bulge S\'ersic index with structural properties
  of bulges. There is a clear and distinct break in these correlations
  at S\'ersic index of $\nb=2$. This break is consistent with a
  picture in which bulges with larger S\'ersic index ($\nb>2$) are
  physically similar to elliptical galaxies, and those with smaller
  S\'ersic index ($\nb<2$) are a different class of
  object. \label{fig:sersic_structure} }
\end{figure} 

In addition, if we restrict the sample to only those galaxies where
the resolution is better than 300~pc, the correlation becomes
stronger. In the improved-resolution sample, we find that only 6\% of
the classical bulges have low S\'ersic index and roughly 9\% of
pseudobulges have high S\'ersic index. If we exclude both S0 galaxies
and those galaxies that are poorly resolved, the correlation improves
still. In this case only 4\% of classical bulges have $\nb<2$.

Exactly at what resolution the use of S\'ersic index becomes
unreliable is difficult to say. Nonetheless, even with the very loose
cut applied here we already detect a difference in
S\'ersic index. As mentioned above, fitting S\'ersic functions to
galaxy light profiles is a very degenerate procedure. If a bulge
diameter approaches the beam width of the data set, clearly using
S\'ersic index to diagnose bulge types would be unreliable in this
scenario. Thus, if bulges are typically $\sim$2~kpc in diameter, then
surveys using SDSS only to measure bulge properties should not extend
beyond $z=0.03$ or a distance of $\sim $120~Mpc, in which a seeing of
1.5 arcseconds would allow for a few resolution elements to sample the
bulge.

We remind the reader that this correlation is an empirical
result. Broadly speaking, the observation that pseudobulges would have
nearly-exponential surface brightness profiles, and thus be more
similar to what is observed in disks, is consistent with the general
observation that pseudobulges are disk-like.  Yet, the physical reason
that such a sharp dividing line in S\'ersic index at $\nb=2$ exists
separating bulges of different morphological types, is not well
understood. Furthermore, the exact distribution of S\'ersic indices
for pseudobulges and classical bulges is hard to establish for
multiple reasons. First, if classical bulges and elliptical galaxies
are truly a single class of object, then ellipticals should be
included in any analysis of surface brightness profiles. Including
early-type galaxies would lead to more galaxies with larger S\'ersic
index \citep{caon94,kfcb,blanton2005}. Secondly, galaxies in which
both a pseudobulge and classical bulge are present would complicate
this analysis. Such systems have been estimated to make up $\sim$10\%
of bulge-disk galaxies \citep{fisherdrory2010}. Thirdly, it is
difficult to compile large samples of unbiased pseudobulge
identification methods that are independent of the S\'ersic
index. Nuclear morphology enabled by the HST archive and S\'ersic
index are the most widely available sources of pseudobulge
detection. It is difficult to obtain, for example, kinematics with
sufficient spatial and spectral resolution on a large number of
galaxies. Also, as we will discuss later using star formation rates
and/or stellar populations is subject to biases in the detected
systems.

The correlations of structural properties with S\'ersic index show a
distinct change at $\nb=2$ \cite{fisherdrory2008,fisherdrory2010}. In
Fig.~\ref{fig:sersic_structure} we show the correlation of bulge
S\'ersic index with half-light radius and effective surface
brightness. Bulges with $\nb>2$ show behavior consistent with that of
E-type galaxies, that is to say a positive correlation between galaxy
size (or luminosity) and S\'ersic index
\citep[e.g.][]{graham96,photplane,kfcb,falcon2011}. Bulges with
$\nb<2$ do not participate in these correlations, and in fact show a
lack of scaling relationships between $\nb$ and other structural
quantities. This is clearly evident in $\re-\nb$ parameter space.

In following sections we will discuss in more detail the correlations
of bulge S\'ersic index with kinematic, interstellar medium, and
stellar population properties of bulges. Bulges with $\nb<2$ are
observed to have higher fractions (and surface density) of gas
\citep{fisher2013}, that is more actively forming stars
\citep{fdf2009,gadotti2009,fisherdrory2010}, and has more disk-like
kinematics \citep{fabricius2014} when compared to bulges with $\nb>2$.
These results, and those in Fig.~\ref{fig:sersic_structure}, suggest
that the S\'ersic index is sensitive to physical differences between
bulge types.

\subsection{Differences in bulge types fundamental plane parameter space}

\begin{figure}
\begin{center}
\includegraphics[width=0.85\textwidth]{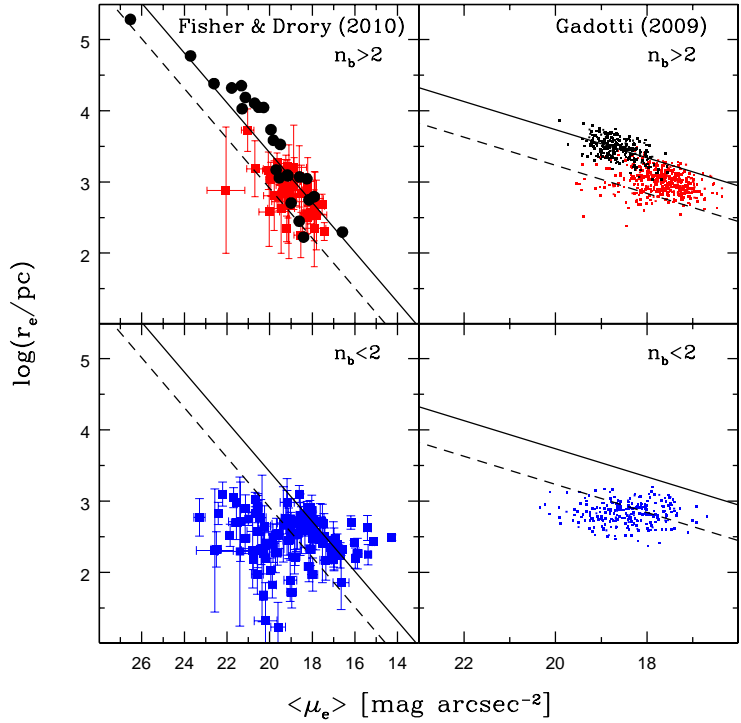}
\end{center}
\caption{ Here we show the relationship of $<\mue> - \re$ for bulges
  (red \& blue squares) and elliptical galaxies (black circles) using
  data from composite profiles of HST/NICMOS, Spitzer 3.6~$\mu$m and
  2MASS data (the magnitude scale is set to match 3.6~$\mu$m scale)
  from \citet{fisherdrory2010} (left), and SDSS $i$
  \citet{gadotti2009} (right). In both cases we show a correlation fit
  to the ellipticals (solid line) and a line set to contain the spread
  in elliptical galaxies (dashed line). The results of these studies
  are essentially consistent, there is a significant population of
  bulges that deviates toward lower surface brightness from this
  projection of the fundamental plane. 
  \label{fig:kormendy} }
\end{figure} 

Elliptical galaxies follow a very well-known set of correlations
between surface brightness, radius, and velocity dispersion, known as
the ``fundamental plane''
\citep{djorgovski1987,faber1989,k77,bbf92}. These relationships are
derived from the assumption that elliptical galaxies are virialized
systems, with small - but significant - deviations corresponding to
variation in mass-to-light ratios and the non-homology of such
galaxies.  Because simulations predict that structural scaling
relations like the fundamental plane are likely to emerge through the
merging processes that form elliptical galaxies through violent
relaxation \citep[e.g.][]{boylan2006}, it would seem reasonable that
if pseudobulges, which are more disky, form significantly differently
than elliptical galaxies and classical bulges they would not
necessarily occupy the same correlation.

There is, however, a danger to using the fundamental plane to identify
bulge types. There is no independent theory that predicts the location
of pseudobulges in these correlations, and there is nothing to say
that in certain projections of fundamental plane correlations
pseudobulges and classical bulges would not overlap. We will
continuously argue throughout this review, there does not seem to be a
single ideal way to identify pseudobulges and classical bulges. A
comprehensive approach that combines multiple indicators of bulge
types is therefore called for.

\citet{carollo1999} shows that the centers of spiral galaxies that
contain pseudobulges have lower surface density than classical
bulges. The location of bulges in projections of the fundamental plane
is studied with larger samples using full bulge-disk decomposition in
\citet{gadotti2009} and also \citet{fisherdrory2010}. Both of these
works find results that are consistent with \citet{carollo1999}, that
is a population of bulges with lower surface brightness than
corresponding elliptical galaxies of similar size or luminosity.

In Fig.~\ref{fig:kormendy} we show the relationship between $<\mue>$
and $\re$ \citep{k77}. The data set we use in this figure is taken from
\citet{fisherdrory2010} (left panels) and \citet{gadotti2009} (right
panels). The data set from \citet{fisherdrory2010} is considerably finer
spatial resolution and uses near-IR data less affected by
variations in mass-to-light ratios and extinction. The
\cite{gadotti2009} sample is a much larger, uniformly selected sample of
nearly $10^3$ galaxies from SDSS, and therefore offers a statistically
sound data set. Both of these studies find essentially the same
result, a significant fraction of bulges deviates toward low surface
brightness. Furthermore, those bulges that
deviate from this relation are much more likely to have low S\'ersic
index. Based on these results, reproduced in Fig.~\ref{fig:kormendy},
it is clear that if a bulge deviates significantly toward low surface
brightness from the \citet{k77} relation, then this is strong evidence
that this bulge is a pseudobulge.

\begin{figure}
\begin{center}
\includegraphics[width=0.85\textwidth]{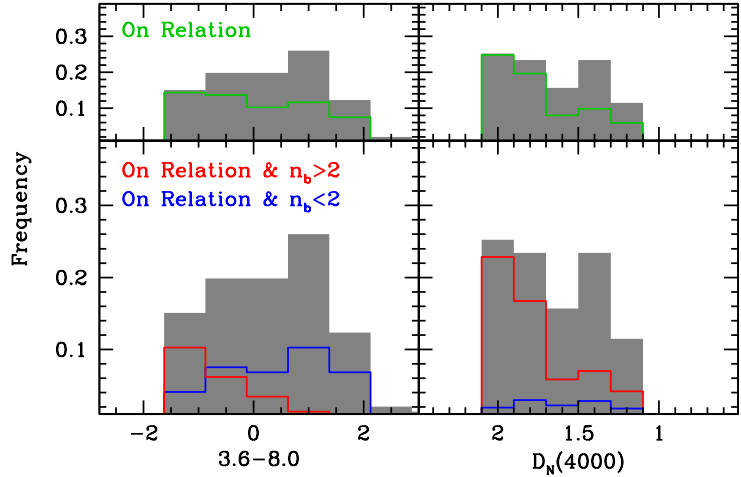}
\end{center}
\caption{ The above figure aims to examine the properties of bulges
  that are consistent with the $<\mue>-\re$ relationship shown in
  Fig.~\ref{fig:kormendy}. 
 The left panels show distribution of $3.6-8.0$
  micron colors from Spitzer IRAC data, measured in
  \citet{fisherdrory2010}. Higher values of $3.6-8.0$ indicate,
  roughly speaking, larger specific star formation rates. The left
  panels show \Dn\ values for bulges (excluding E galaxies)
  from \citet{gadotti2009}. Smaller values of \Dn\ indicate
  younger populatios. Note that we have inverted the x-axis of
  \Dn\  so that in all panels younger, higher star forming
  bulges are on the right side of the panel. The grey shaded region
  shows the distribution for the entire sample. The green line
  represents all those bulges that are consistent with the
  $<\mue>-\re$ relationship. The blue line is those bulges that are consistent with the
  $<\mue>-\re$ and have $\nb<2$. The red line shows the distribution
  for bulges consistent with  $<\mue>-\re$ and have $\nb<2$. 
  \label{fig:re_mue_sf_test}}
\end{figure} 

Identifying bulges as classical bulges because they are consistent
with the $<\mue> - \re$ relationship, however, is less robust.
\cite{gadotti2009} marks bulges contained within the spread of the
\citet{k77} relation as classical bulges. They argue that at least in
this parameter space, these bulges are structurally similar to
elliptical galaxies.  This makes the assumption that other physical
processes cannot make a bulge with similar values of surface
brightness and size. Absent a result from simulations, we cannot know
if that assumption is true.

We can look at the properties of those bulges that are consistent with
the \citet{k77} relation to determine how homogenous a class they are.
In Fig.~\ref{fig:re_mue_sf_test} we show the distribution of
$3.6-8.0~\mu$m color \citep{fisherdrory2010} and \Dn\
\citep{gadotti2009} for the bulges that are consistent with the
\citet{k77} relation. Larger values of $3.6-8.0~\mu$m color imply more
active star formation per unit stellar mass. Smaller values of
\Dn\ imply younger stellar populations; for display purposes we
plot the \Dn\ values in reverse order, so in both panels
younger, more star forming systems are on the right side of the
panel. In both samples it is clear that selecting bulges only by the
location in $<\mue> - \re$ parameter space does not uniquely separate
bulges. In the bottom panel we show the combination of using both the
\citet{k77} relation and $\nb$ as selection criteria for bulge types.
In the \citet{fisherdrory2010} sample this more cleanly identifies
classical bulges as non-star forming systems.

In summary, using the fundamental plane as bulge type diagnostic
carries certain caveats. If a bulge significantly deviates toward
lower surface brightness from the \citet{k77} relationship between
$<\mue>$ and $\re$, then this is strong evidence that that bulge is a
pseudobulge, based on studies of its star formation rate, S\'ersic
index, and nuclear morphology. However, if a bulge has parameters
consistent with the fundamental plane, from an empirical
point-of-view we cannot say what type of bulge this is. For example,
if the aim of a study is to isolate a sample of bulge-disk galaxies
that resemble M~31 (a prototypical classical bulge), then using
$<\mue> - \re$ alone is clearly insufficient, and as we show in
Fig.~\ref{fig:re_mue_sf_test} this method selects a number of star
forming bulges. Also, \citet{fisherdrory2010} show that a number of
bulges that are consistent with the \citet{k77} relationship have
nuclear morphology that, unlike M~31, resembles a disk.

\section{The interstellar medium and stellar populations of
  pseudobulges and classical bulges}

Historic work concluded that bulges are uniformly old and devoid of
star formation \citep[e.g.][]{whitford1978}. This led to the widely
held view that all bulges are old and inactive. This turns out to be
true for some bulges, but it is not universally true by any means.
For example, in the prototypical classical bulge of M~31, the dust SED
is consistent with being completely heated by the old stars, and shows
no evidence for new star formation \citep{draine2014}, and also the
stellar populations indicate a uniformly old population of stars, with
mean ages above 12~Gyr \citep{saglia2010}. However, work in the last
15-20 years shows that many bulges contain cold gas, actively form
stars and can have short mass doubling times, and often have
intermediate-to-young light-weighted stellar ages.

\citet{peletier1996} show that some bulges are indeed quite blue, and
that in general bulges have similar optical colors as the surrounding
disk. Similarly, \citet{regan2001bima} finds using interferometric
observations of CO(1-0), that some bulges are as gas rich (from
$L_{\rm CO}$-to-$L_{\rm K}$ ratios) as the associated outer disk. In
the past 10 years data from Spitzer Space Telescope, GALEX UV
telescope, and CO interferometry from BIMA, OVRO, CARMA \& PdBI have
greatly improved our ability to measure star formation rates in
bulges. We can now robustly say that specific star formation rates and
gas fractions in the bulge region of nearby galaxies are often very
high
\citep{sheth2005,jogee2005,fisher2006,fdf2009,fisherdrory2011,fisher2013}. Also,
bulges can contain young stellar populations
\citep{gadotti2001,macarthur2004,peletier2007,ganda2007}. See also
\citet{kk04} for a review.

From a physical perspective it makes sense that pseudobulges would be
systematically younger with more active star formation than classical
bulges. The present model is that classical bulges formed in the
early Universe, either through merging
\citep{aguerri2001,robertson2006} or as the result of clumpy disk
instabilites \citep{noguchi1999,elmegreen2008}. The former become less
frequent, and the latter are extremely rare below $z\sim
1$. Conversely, galaxies with pseudobulges either did not experience
these processes, or they were signifcantly less pronounced, the
resulting galaxy was able to evolve secularly for long periods of time
and still does. Some of them still contain significant amount of gas
to fuel internal evolution of the bulge. Also, the presence of a
classical bulge may in fact stabilize a galaxy against star formation,
and especially the secular inflow of gas \citep{martig2009}. This
process known as ``morphogical quenching'' may act to reinforce a
correlation with bulge structural properties and bulge star formation
rates.

Before going on, we must point out a simple, yet critical, caveat to
using stellar populations, star formation rates, and gas fractions to
identifying pseudobulges. Gas stripping by cluster environments
\citep[as described by][]{kenney2004} can shut down star formation in
a galaxy. If such a galaxy had previously formed a pseudobulge, that
bulge would quickly appear inactive and old.  Also, simulations show
that pseudobulges can form in dissipationless systems
\citep{debattista2004}. It is therefore important that one should not
use the absence of star formation alone as a reason to suggest a
galaxy does not contain a pseudobulge.

To be clear, when we refer to ``bulge'' star formation rates and gas
masses what we really mean is the star formation rate (or gas mass)
inside the region of the galaxy where the bulge dominates the
light. Cold gas, and thus star formation, happens in a thin disk
\citep{garcia1999} of scale height of $\leq 100$ parsecs. Bulge-disk
decompositions, however, do not typically consider the thickness of
the bulge.  Indeed, the thickness of pseudobulges is very poorly
constrained. Some are likely very thin \citep[as argued by][]{k93},
however, given the common presence of resonant phenomena it is likely
that many are thickened. If the goal is to understand how properties
of the bulge evolve, however, then comparing the entire mass (or
luminosity) of the bulge stars to the entire rate of star formation in
a bulge seems appropriate.

\subsection{A brief aside on measuring star formation rates in bulges}

The measurement of star formation rates in galaxies, $SFR$, is
typically done by means of a tracer of the amount of young stars
present. This field has greatly advanced in the past decade
\citet{kennicutt98araa,calzetti2007,kennicutt2009,leroy2012,kennicuttevans2012}.
Because the emission from O and B stars heavily dominates the UV
spectral range, it is straightforward to argue that $SFR \propto
L_{\rm UV}$. The calibration of such a relationship can be found in
\citet{salim2007}.  For bulges, data from the GALEX UV space telescope
is well suited to resolve $\sim 1$~kpc in galaxies within 40~Mpc. An
alternative approach is to use emission from HII regions, typically
this is done using the H$\alpha$ flux, assuming in this case that
$SFR\propto L_{H\alpha}$.

A difficulty to estimating the emission from young stars is that dust
absorbes UV/optical emission.  This is especially important for galaxy
centers (i.e.~bulge regions), which experience more extinction
\citep{peletier1999,macarthur2004}.  In fact, we know from studies of
our own galaxy that star formation can occur in heavily obscured
regions \citep[for review][]{evans1999,kennicuttevans2012}, and
therefore much of the light may be missed in optical observing
campaigns.

One way to overcome the effects of extinction would be to measure the
flux of a Hydrogen emission line in the near-infrared range
(e.g.~Pa~$\alpha$ emission). However, such measurements can be
difficult to make, and are often low signal-to-noise. Alternatively,
data from {\it Spitzer Space Telescope} allows us to directly probe the
re-radiation in the infrared of the energy absorbed by the dust in the
UV/optical, for example using emission at 24~$\mu$m
\citep{calzetti2007}. A common approach in the current literature is
to combine different star formation tracers
\citep[e.g.][]{kennicutt2009} to account for both the unobscured star
formation (traced by UV or H$\alpha$) and the obscured star formation
(traced by infrared emission).

The 8~$\mu$m emission is dominated by polycyclic aromatic hydrocarbons
(often called PAHs). At present, and with respect to measuring the
properties of bulges, a significant advantage of 8~$\mu$m maps
available from Spitzer is that they have significantly finer spatial
resolution (beam size of Spitzer IRAC 8~$\mu$m data is $\sim 2$
arcsec, roughly 3 times better than 24~$\mu$m maps with
MIPS). However, flux from the 8~$\mu$m emission is not reliable as a
direct, one-to-one, indicator of the star formation
rate. \cite{calzetti2007} show that the correlation between
continuum-corrected 8~$\mu$m flux and Pa~$\alpha$ flux depends on
both environment and metallicity. In light of this, we limit our use of
PAH emission to mostly an on/off metric of activity, separating star
forming bulges (bright 8~$\mu$ emitting bulges) from non-star forming
systems.

A second issue in measuring star formation in galaxy bulges is the
contamination of the metrics of star formation by old stars. Old
stellar populations make a measurable contribution of light at UV
wavelengths and lead to an overestimate of the star formation rate
\citep[e.g.][]{cortese2008}. Also, old stars can heat dust and thereby
increase the 24~$\mu$m flux. These problems are especially pronounced
in bulges where the surface density is very high. In this case the
flux in tracers used to probe star formation is actually a combination
of contributions from old and young stars. \citet{leroy2012} study
this in galaxy disks by modelling the diffuse emission. They find that
typically roughly 20\% of the emission at 24~$\mu$m can be atributted
to evolved stellar populations.  \citet{fisher2013} investigate bulges
specifically and find similarly that in typical star forming bulges
the star formation rate is decreased by roughly 20\% when accounting
for old stellar populations. Both \citet{leroy2012} and
\citet{fisher2013} show that this effect is stronger in regions of low
star formation. \citet{fisher2013} also shows, as expected, that this
effect is more pronounced when the surface density of star light is
higher.  For example, in the bulge of M~31, \citet{draine2014} find
that essentially all of the dust emission is accounted for by heating
by stellar populations.

\subsection{Active star formation and more gas is strongly correlated
  with bulge types}

\begin{figure}
\begin{center}
\includegraphics[width=0.6\textwidth]{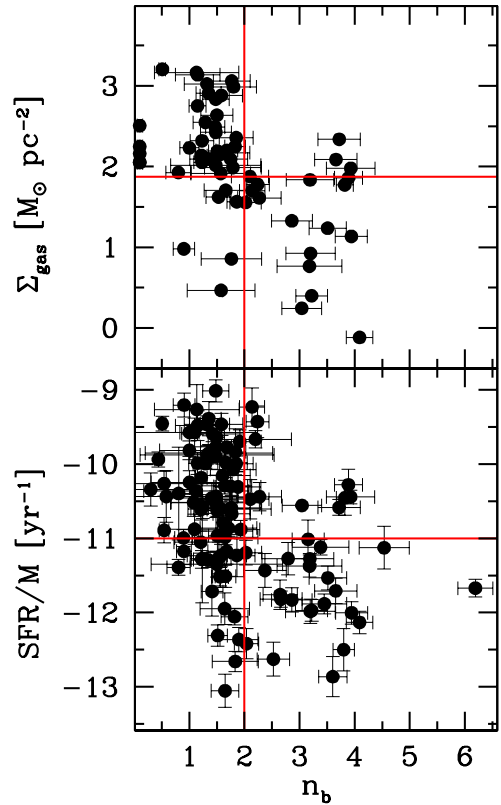}
\end{center}
\caption{The above figure compares the specific star formation rate
  (bottom panel) and gas surface density (top panel) of bulges to the
  bulge S\'ersic index. Data are taken from
  \citet{fdf2009,fisherdrory2011,fisher2013}. The vertical lines
  indicate the commonly used pseudo-classical bulge dividing line of
  $\nb=2$, the vertical lines are set to guide the eye for $SFR/M=
  10^{-11}$~yr and
  $\Sigma_{\rm gas}=75$~M$_{\odot}$~pc$^{-2}$. \label{fig:sfr_n}}
\end{figure}

Though there is a long history of evidence that many bulges are
actively forming stars, and that star formation is likely
significantly altering the stellar structure of a bulge \citep[for
  review, and outlined above]{kk04}[], direct comparisons to bulge
classifications began mostly recently.  A general summary is that if a
bulge is star forming, statistically speaking it likely has other
pseudobulge properties (e.g.~$\nb<2$, disky nuclear
morphology). Conversely, if a bulge is not star forming it can have a
large mix of properties, consistent with the discussion above. Also, a
small subset of galaxies have star forming centers, but quiecent
disks; these systems also have large bulge S\'ersic index. A plausible
scenario to explain these systems could be the recent accretion of a
satellite directly into the galaxy center \citep[e.g.][]{aguerri2001}.

\citet{fisher2006} uses data from Spitzer Space Telecope and archival
HST data to directly compare the morphological diagnosis of bulge
types to the $3.6-8.0$~$\mu$m color profiles of galaxies with
pseudobulges and those with classical bulges. In this case,
$3.6-8.0$~$\mu$m color is a very rough proxy for specific star
formation rate (star formation rate divided per unit stellar mass).
They find that in galaxies with classical bulges, the color of the
disk indicates active star formation, however there is a sharp break
near $\sim 1$~kpc where the color profile transistions to a
non-starforming bulge. In contrast, there is no such transition in
galaxies with pseudobulges. The pseudobulge is forming stars similarly
to the outer disk. \citet{fisherdrory2010} follow this up by
calculating the $3.6-8.0$~$\mu$m color for $\sim$180 galaxies, and
study other indicators of bulge type (morphology, S\'ersic index, and
$\mue-\re$). They find that if a bulge has mid-IR colors satisfying
$3.6-8.0>0$, then that bulge has properties
that resemble a pseudobulge (eg. low $n_b$). 

In Fig.~\ref{fig:sfr_n}, we compare the specific star formation rate
and gas surface density of bulges to the bulge S\'ersic index, using
data from \citet{fdf2009}, \citet{fisherdrory2011}, and
\citet{fisher2013}. The results here re-iterate the results of these
papers. In both panels, it is clear that active star formation and
high surface densities of gas are exclusively found in bulges with low
S\'ersic index. It is worth pointing out that there is no {\em a
  priori} reason that the bulge S\'ersic index would correlate with
the bulge gas density; a similar correlation is recovered if one
measures bulge gas density with a fixed radius (e.g.~500~pc) and if
one uses the bulge radius as done here. These correlations imply that
the seperation of bulge types is likely tied to a physical
distinction. The results in Fig.~\ref{fig:sfr_n} continue to motivate
that the separation of bulges into at least two categories is
informative to the physics of galaxy evolution.

Using the star formation rate (or gas surface density) of a bulge
alone to identify it as a pseudobulge or classical bulge is,
statistically speaking, somewhat ambiguous. 8\% of bulges that have
active star formation (defined as $SFR/M>10^{-11}$~Gyr$^{-1}$) also
have $\nb$ that is significanlty larger (considering error bars) than
$\nb=2$. A similar result is true when considering $\Sigma_{\rm
  gas}>75$~M$_{\odot}$~pc$^{-2}$. Therefore, if one discovers that a
bulge has a very active gas rich center, this is strong evidence for
that bulge being a pseudobulge. However, it is clear that when the
star formation or gas density is low, one should not infer the bulge
type.  We recommend using star formation as a ``second tier'' method
for identifying pseudobulges and classical bulges. For example if
other metrics give ambiguous results but the bulge is very actively
forming stars one could then conclude the bulge is a pseudobulge.

\subsection{Stellar population indicators and bulge types}

Stellar population indicators in bulges show a wide range in
properties \citep[for a brief review see][]{peletier2008}. The topic
of stellar populations is quite broad with a large variety of
techiques and results that could easily fill its own review. We will
concentrate on those results in which correlations, or the notable
lack thereof, are relevant as diagnostics of bulge type.  There is no
set of stellar population parameters that is typical of a bulge. As
mentioned before, an overwhelming majority of studies shows that the
historic assumption that all bulges are uniformly old is simply not
supported by the data
\citep[e.g.][]{dejong1996,peletier1996,carollo2001,proctor2002,moorthy05}.

\begin{figure}
\begin{center}
\includegraphics[width=0.9\textwidth]{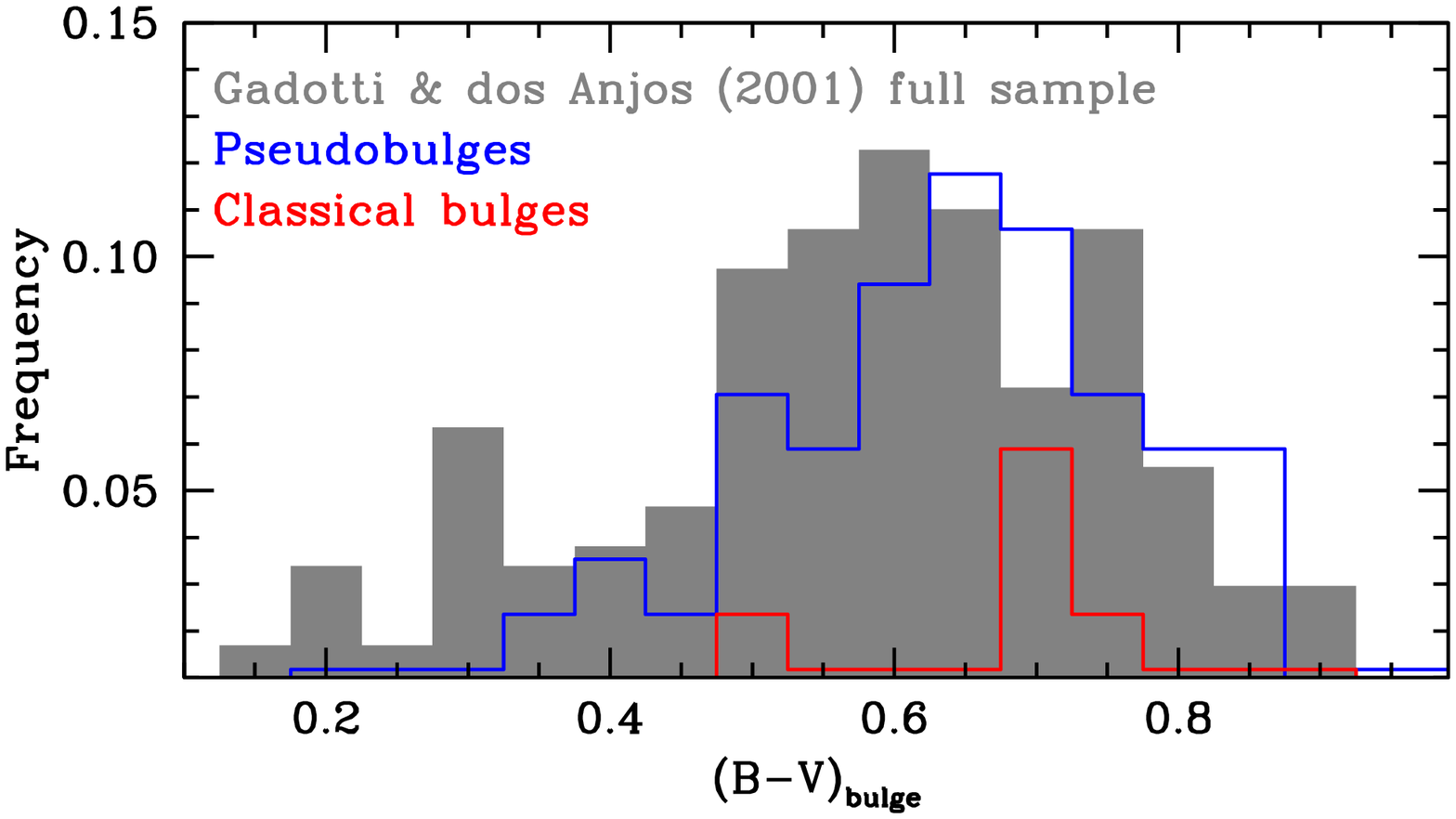}
\end{center}
\caption{Distribution of $B-V$ for bulges in the \citet{gadotti2001}
  sample of galaxies (shaded region). The blue line represents those
  bulges that are identified as pseudobulges and the red line
  represents those that are classified as classical bulges (by
  combining S\'ersic index, nuclear morphology and the Kormendy
  relationship). Though classical bulges are rarely found to be blue,
  pseudobulges very often have red optical
  colors.\label{fig:bmv_bulge}}

\begin{center}
\includegraphics[width=0.9\textwidth]{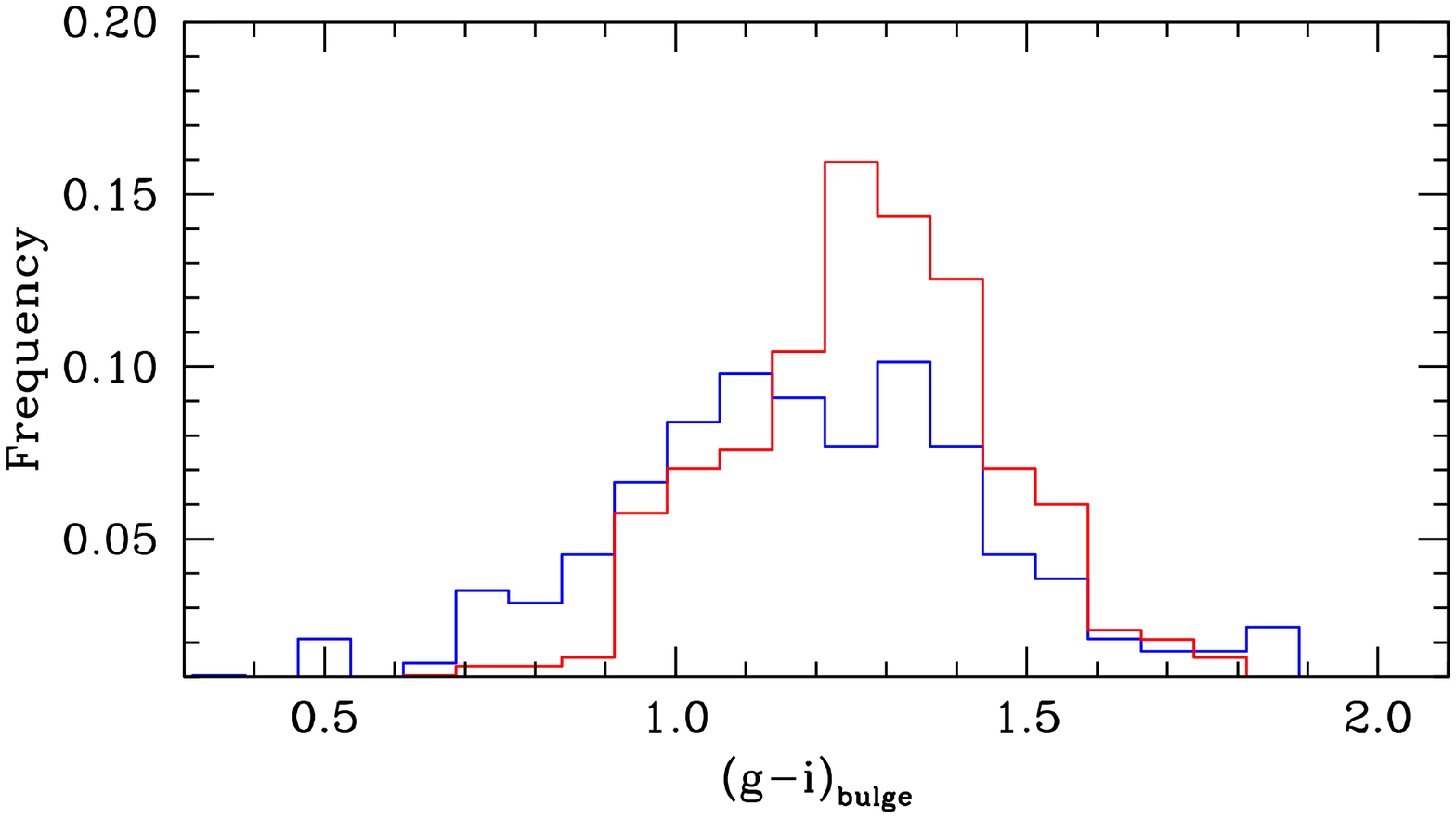}
\end{center}
\caption{Distribution of $g-i$ for bulges in the \citet{gadotti2009}
  sample of galaxies. The blue line represents those
  bulges that are identified as pseudobulges and the red line
  represents those that are classified as classical bulges (by
  combining S\'ersic index,and the Kormendy
  relationship). \label{fig:gmi_bulge}}
\end{figure}

There has been mixed evidence that optical color can be used as a
means of identifying pseudobulges. Early results were promising. For
example, \citet{peletier1996} found a large spread in ages of bulges,
and that average stellar age of bulges correlates with that of disks
(young bulges are in young disks). This was confirmed in a much larger
samples by \citet{gadotti2001,macarthur2004}. \citet{carollo2001} find
that the average $V-H$ color of exponential bulges with disky nuclear
morphology (i.e.\ pseudobulges) is bluer than that of $r^{1/4}$
bulges.

Studies of the color of larger samples of bulges suggest that a single
broadband color using optical or near infrared filters do not
correlate strongly enough with other indicators of pseudobulges for
reliable use.  In Fig.~\ref{fig:bmv_bulge} we show the distribution
of bulge colors from \citet{gadotti2001} (grey shaded area). We also
cross-reference the sample from Fig.~\ref{fig:sersic} against 3 papers
which contain samples of the same bulge color
\citep{gadotti2001,mollenhoff2004,fisher2013}\footnote{Fisher et
  al.~(2013) use SDSS $g-r$, which we convert to $B-V$ via
  \citet{smithetal2002} transformations.}. Pseudobulges are identified
as bulges which have any of the following: $\nb<2$, nuclear morphology
that resembles a disk, and/or low surface brightness outliers from
$\mue-\re$ relation. The \citet{gadotti2001} sample is shown to
ensure that our bulge classification sample is not significantly
biased. The distribution of classical bulges clearly skews to the
redder colors, similar to \citet{carollo2001}. Blue bulges are far
more likely to be pseudobulges. However unlike the previous methods of
identifying pseudobulges there is not a significant range in this
parameter over which classical bulges are not found. 

The lack of a strong correlation between bulge color and type is likely
not due to sample selection. \citet{gadotti2009} finds a similar
result using $g-i$ colors. In Fig.~\ref{fig:gmi_bulge} we show the
distribution of bulge colors for the 670 bulge-disk galaxies from
\citet{gadotti2009}. \citet{fernandexlorenzo2014} find a similar
result with 189 galaxies, albeit the sample is biased only to include
isolated galaxies. 

\citet{gadotti2009} also compare the stellar populations tracer \Dn\
\citep{kauffmann2003} to bulge types (determined from bulge-disk
decompositions). The break in the optical spectrum which occurs at
4000~\AA  is smaller for younger stellar populations \citep[for
description see]{bruzual1983,kauffmann2003}, and is a good identifier
of young or bursty populations.  \citet{gadotti2009} find indeed that
pseudobulges have on average smaller values of \Dn\ and therefore
pseudobulges are more likely to be young, but again there is not a
significant range that isolates one type of bulge.

Taken all together, these results suggest that there tends to be a
preference for pseudobulges to be blue {\em on average} compared to
classical bulges. This particular subject could benefit from a work
with both a well-defined and large sample of galaxies that is well
resolved.  However, based on the data that presently exists, optical
color on its own is not a reliable indicator of bulge-type.

Similar to the results from optical colors, studies of bulge ages
using more robust techniques such as absorption line indices or
spectroscopic synthesis, return mixed results (see
\citealp{renzini2006} and references therein for a discussion of these
techniques). \citet{proctor2002} shows that bulges are younger on
average and have fewer metals than early type galaxies. Both
\citet{moorthy05} and \citet{thomas2006} find a wide spread in ages,
and that many bulges in later type (Sb-Sbc) galaxies are quite old.
\citet{macarthur2009} find, similarly, that the fraction of mass in
bulges that was formed in the past gigayear is quite small.
\citet{zhao2012} uses the Sloan Digital Sky Survey to measure the
stellar populations of bulges in a sample of 75 isolated
galaxies. Bulge types are diagnosed using both S\'ersic index and the
$\mue-\re$ relation, and they find that on average pseudobulges have
more prolonged star formation than classicals. \citet{zhao2012} find
no classical bulges that are younger than $\sim 6$~Gyr (mass weighted
age), conversely roughly 30\% of pseudobulges are found to be younger
than this. However, the average age difference between the two
populations of bulges is not very large.

Differences in stellar population indicators do exist between the
bulges of different types. A particularly significant difference is
found in the absorption line indices of bulges
\citep{peletier2007,ganda2007}. It is well known that for elliptical
galaxies the Mg$_2$ line index correlates well with velocity disperion
\citep[e.g.][]{bbf92}.  \citet{peletier2007} and \citet{ganda2007}
show that many bulges fall below this relation, especially those
bulges with low velocity disperion centers and/or those bulges in
late-type galaxies.

\begin{figure}
\begin{center}
\includegraphics[width=0.8\textwidth]{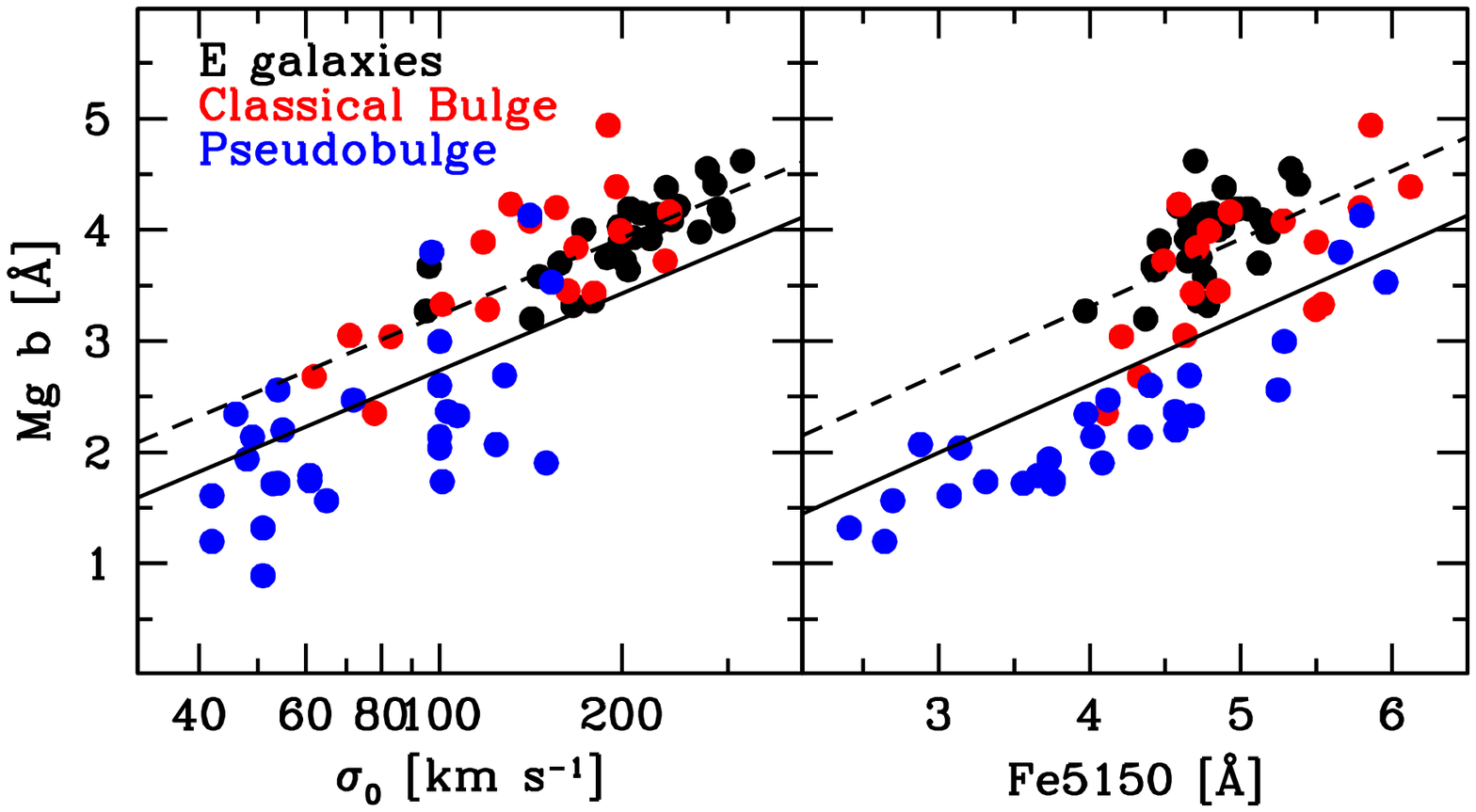}
\end{center}
\caption{ The $Mg~b-\sigma$ relationship for elliptical galaxies and
  bulges. We have restricted the control sample (black dots) to only
  the E galaxies from the SAURON sample to ensure that no pseudobulge
  galaxies are included.  We take the bulge values from
  \citet{peletier2007} and \citet{ganda2007}. Quantities for
  pseudobulges (identified with nuclear morphology, S\'ersic index and
the $\mue-\re$ relation) are plotted as blue points, and the red
points represent classical bulges. The dashed line is offset 0.5 $\AA$
below the best fit relation (solid) line. Only pseudobulges are found
below this line. \label{fig:mg}}
\end{figure}

In Fig.~\ref{fig:mg} we show that a strong connection exists between
bulge type and absorption line indices, specifically Mg~b and
Fe5150. For this figure we show the central values of pseudobulges,
classical bulges and elliptical galaxies taken from a sample combining
data from \citet{ganda2007,peletier2007,kuntschner2010}. We have
classified bulges using bulge S\'ersic index and bulge morphology. No
classical bulge or elliptical galaxy has Mg~b$<$2.35~\AA, conversely
over 2/3 of pseudobulges have lower values. (See
\citealp{ganda2007,peletier2007,kuntschner2010} for a discussion of
these Lick indices.) Similarly, the lowest value of Fe5150 in
classical bulges and elliptical galaxies is 3.97~\AA\ whereas roughly
50\% of pseudobulges are found below this limit. As we show in
Fig.~\ref{fig:mg}, the Lick indices become particularly powerful when
combined with the velocity dispersion. In each panel the dashed line
is a best fit relationship to the E galaxies and classical bulges, the
solid line represents a relation with the same slope, yet offset down
in Mg~b such that it seperates pseudobulges and classical bulges.

Based on this data set a bulge is a pseudobulge if it meets any of the
following criteria: \\
\noindent 1. Fe5150 $< 3.95~\AA$, \\
\noindent 2. Mg~b$<2.35~\AA$, \\
\noindent 3. $\Delta Mg~b<0.7~\AA$ compared to the $Mg-\sigma$
correlation, \\
\noindent 4. $\Delta
Mg~b<0.7~\AA$ compared to the $Mg-Fe$ relation. 

All low Mg~b outliers to the $Mg-\sigma$ relation are also
outliers to the $Mg-Fe$ relation, but the reverse is not
true. Conversely, there is more spread in the classical bulges in
$Mg-Fe$. We also stress that because the sample of bulges includes a
number of old S0 galaxies from the SAURON survey, using these Lick
indices to identify pseudobulges and classical bulges appears to be
robust against age. So it seems that using both of these relationships
together would be a powerful tool for identifying pseudobulges,
especially in the near future in which surveys such as SAMI and MANGA
will measure absorbtion line strengths for large numbers of galaxies.

\section{Identifying Pseudobulges and Classical Bulges with Kinematic Properties}

Kinematic measurements of bulges provided some of the earliest
evidence for the dichotomous nature of bulges. \citet{k82} points out
that some bulges in barred disks are kinematically more similar to
disks than those in unbarred disks. This kinematic similarity is
indicated by the ratio of peak rotation velocity to bulge velocity
dispersion, which is taken as a proxy of the ratio of
``ordered-to-random motions.''  Indeed, use of the $V/\sigma -
\epsilon$ parameter space can distinguish pseudobulges from classical
bulges, as shown by \citep{k93,kk04,kormendyfisher2008}. However,
these studies rely on very small samples, fewer than 20 bulge-disk
galaxies, and are thus difficult to control. At present, it is safe to
say that bulges with values well above the ``oblate line'' in the
$V/\sigma - \epsilon$ parameter space are considered to be rotating
bulges, and thus from a theoretical perspective would be
``pseudobulges'', but it is difficult to estimate empirically, using
currently available data, how often a bulge that has low S\'ersic
index would also be found in the ``disk'' region of the $V/\sigma -
\epsilon$ diagram.

\begin{figure}
\begin{center}
\includegraphics[width=0.8\textwidth]{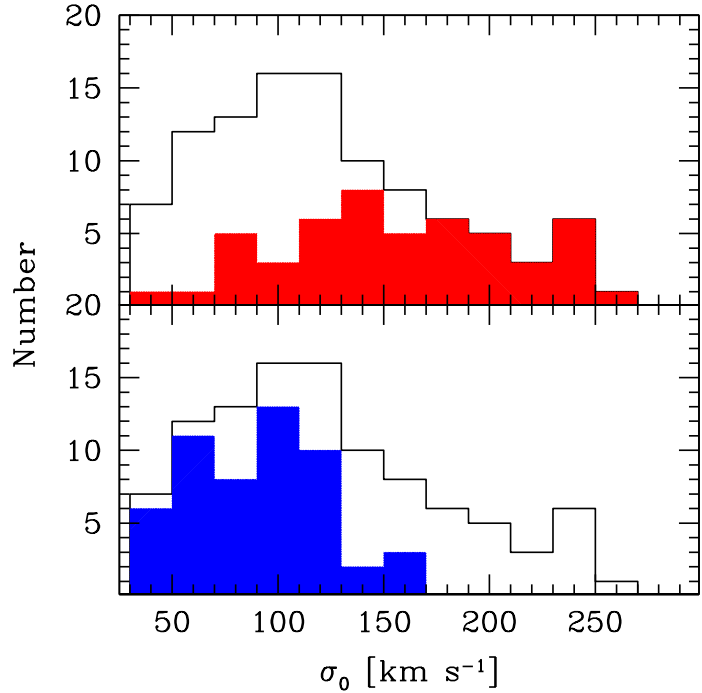}
\end{center}
\caption{ Distribution of central velocity dispersions for galaxies
  with published $\sigma_0$ in the sample from
  Fig.~\ref{fig:sersic}.  As before, pseudobulges are indicated by
  blue shaded region, and classical bulges by red. There is clearly
  significant overlap between the two samples.   \label{fig:sig}}
\end{figure}

Central velocity dispersion alone does not completely separate bulge
types. The distribution of $\sigma_0$ for a sample of $\sim 100$ S0-Sc
galaxies is shown in Fig.~\ref{fig:sig}. To construct the kinematic
sample we use data from published sources that have sufficient
velocity resolution to measure the central dispersion of bulges
($\sigma_{bulge} \geq 50$~km~s$^{-1}$), and also have sufficient
spatial resolution to isolate the absorption line kinematics in the
bulge region ($r_{bulge}\sim 1$~kpc) that also have available
bulge-disk decompositions from the sample used in
Fig.~\ref{fig:sersic} of this review. We use velocity dispersions from
\citet{heraudeau1999,barth2002,ganda2007,kuntschner2010,fabricius2012}. In
this comparison we identify pseudobulges as having $n_b<2$ or
prominent disk-like nuclear morphology as described
earlier. \citet{zhao2012} finds that the distribution of central
velocity dispersions of pseudobulges is essentially the same when
they identify pseudobulges using S\'ersic index or with the Kormendy
relation.  On average, pseudobulges have lower central
velocity dispersion than classical bulges ($<\sigma_0>_{pseudo} \sim
90$~km~s$^{-1}$, compared to $\sim 160$~km~s$^{-1}$ for classical
bulges).  There is a strong decline in the number of pseudobulges with
$\sigma > 130$~km~s$^{-1}$. However, roughly $\sim 1/3$ of the
classical bulges in this sample have $\sigma < 130$~km~s$^{-1}$. It is
for this reason $\sigma_0$ alone cannot be used to statistically
isolate all pseudobulges from classical bulges. When a bulge has
particularly high velocity dispersion ($\sigma > 130$~km~s$^{-1}$)
then it is most likely a classical bulge.

A large sample of uniform measurements of velocity dispersion
preferably with integral field spectroscopic measurements in
bulge-disk galaxies would have significant value in understanding
pseudobulge and classical bulge properties, nonetheless the result in
Fig.~\ref{fig:sig} does not appear to depend on the sample. Both
\citet{fabricius2012} and \citet{zhao2012} find essentially the same
result that we report here. 

\citet{kk04} shows that bulges that are low-$\sigma$ outliers to the
\citet{faber1976} relation between bulge magnitude and velocity
dispersion, are likely pseudobulges. However, there is a significant
amount of spread in this correlation, and similar to the $\mu_e-r_e$
if a bulge is co-located in parameter space with this relationship it
does not mean the bulge is a classical bulge.
\begin{figure}
\begin{center}
\includegraphics[width=0.9\textwidth]{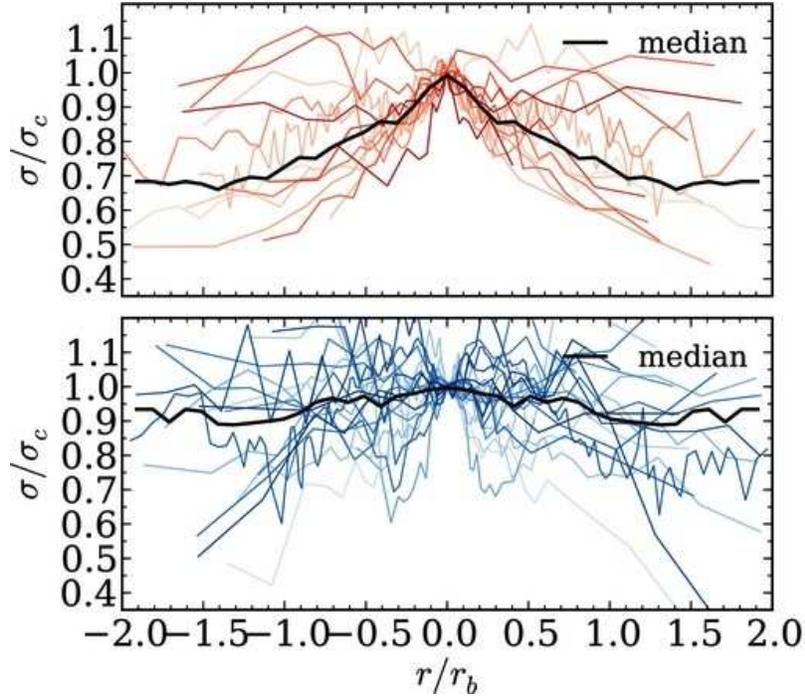}
\end{center}
\caption{ Here we re-plot a result from \citet{fabricius2012} that
  shows how the radial profile of the velocity dispersion in
  pseudobulges is much flatter than that found in galaxies with
  classical bulges. \label{fig:sigprof}}
\end{figure}

Significant correlations between bulge type and the radial structure
of kinematics have been seen by a number of authors
\citep[e.g.~][]{falcon2006,comeron2008,fabricius2012}. In
Fig.~\ref{fig:sigprof} we show the basic result (in this case taken
from \citealp{fabricius2012}) that galaxies with classical bulges have
centrally peaking velocity dispersion profiles, where galaxies with
pseudobulges do not. For this result, \citet{fabricius2012} identifies
pseudobulges using S\'ersic index and bulge morphology. This is
consistent with the overall picture of classical bulges and
pseudobulges. In this case a classical bulge is considered to be a
separate component from the disk, and the classical bulge is
dynamically hotter than the disk.  In the center of the galaxy the
classical bulge dominates the light and the measured kinematics. At
large radius the disk dominates the light, and measured kinematics
have lower dispersion. The intermediate radii show the transition
between these two regimes. Pseudobulges, conversely do not have a
hotter separate component, they are often thought of simply as high
surface density centers of disks, therefore kinematically they do not
break from the behavior of the disk. \citet{fabricius2012} quantifies
the kinematic profile shape with the logarithmic derivative
d~$\log(\sigma)$/d~$\log(r)$. The logarithmic derivative correlates
well with bulge type.  Galaxies with classical bulges ($n_b>2$ and
E-type morphology) have more negative values (i.e.~more centrally
peaking $\sigma(r)$).

\citet{peletier2008} notes that all the bulges with central velocity
dispersion minima in the samples of \cite{ganda2007} and
\cite{peletier2007} also are low-Mg~b outliers (as described
above). \cite{comeron2008} studies the properties of so-called
$\sigma$-drop galaxies (galaxies with a central minimum in velocity
dispersion). They find that dusty structures that would, in this review,
be classified as indicative of pseudobulges are very common in these
galaxies. They also find a higher fraction of circumnuclear star
formation in $\sigma$-drop galaxies. 

\citet{fabricius2012} shows that combining $V/\sigma$ with metrics of
the profile shape can be very powerful for identifying
pseudobulges. Galaxies with classical bulges have low values of
$V/\sigma$ and central cuspy surface brightness profiles.  Essentially
the result is physically sound; if a bulge is dominated by dispersion
and has a higher dispersion to the surrounding disk then it is almost
always a classical bulge. Conversely, pseudobulges are not found in
the same region of parameter space. \citet{fabricius2012} finds that
outliers to this rule tend to be galaxies that in line-of-sight
velocity distributions that these galaxies have multiple kinematic
components that are affecting the measurement of the shape of the
velocity profile. The drawback to this method is that it
requires sufficient velocity resolution to measure the kinematics of
the disk, and therefore may be inaccessible to surveys such as MANGA
and SAMI. 

\section{Composite Pseudo-Classical Bulges} 

Assuming that galaxies either have only a pseudobulge or a classical
bulge is most likely an oversimplification.  Bulges that consist of
both a thin, starforming pseudobulge and a hot-passive classical bulge
are very likely present in some galaxies. There has been very little
work done on composite bulges. This is definitely an area that could
use more work, though results, by the nature of the problem, are
likely to be difficult to interpret.

\citet{fisherdrory2010} argue that scaling relations can be used to
identify some mixed-case bulges. Bulges that are high mass or high
surface brightness outliers from fundamental plane scaling
relationships are likely to be composite. In these systems, a
classical bulge is assumed to be on a scaling relation, for example
the $\mu_e-r_e$ correlation of E galaxies. The pseudobulge component
increases the mass, without strongly affecting the value for the
effective radius. They use models to show that in the limit that the
mass of the classical component is larger than that of the
pseudo-component this is true. \citet{fisherdrory2010} find that
bulges that are co-located in fundamental plane parameter space with
models of composite pseudo-classical bulges have lower specific SFR
than the median pseudobulge, and also have $n_b \sim 1.8-2.1$. They
show by modeling that adding a high S\'ersic index bulge to a low
S\'ersic index pseudobulge tends to produce an intermediate range
$n_b$. \citet{fisherdrory2011} use these results to estimate that
roughly 10-20\% of bulges in the local 11~Mpc, fit this
description. This is only a rough estimate. Much more work is needed
to truly get a robust estimate of the frequency of composite bulge
systems. 

\citet{erwin2015} uses stellar kinematics to model the internal
structures of several examples of galaxies which contain both a
pseudobulge and classical bulge. These models generally find a small
compact structure which is referred to as a classical bulge, with a
diffuse structure around it that has dynamics that are more consistent
with disks, which they call the pseudobulge. 

The take away is that the presence of a pseudobulge in a galaxy does
not necessarily imply that there is not an old, dynamically hot
component of stars within that system. In the future, as integral
field spectroscopy becomes more common, dynamical modelling which
places estimates on the maximum fractional mass of a hot stellar
component in pseudobulges of ranging properties ($B/T$,
$SFR/M_{star}$, $n_b$, etc.) may prove very useful.

\section{Summary}

In this review we have highlighted a number of observed properties
that mark empirical differences between classical bulges and
pseudobulges. We certainly do not always understand the underlying
physical reason for these observed differences. For example, why
$n_b\sim 2$ seems to be such a good dividing line between bulge types
is not clearly understood. An alternate approach is to base diagnostic
methods on physically motivated arguments (such as an assumption on
the star formation history, or the structural properties). However,
physically motivated arguments can be specious, especially when we
consider that theoretical understanding of bulge formation is
incomplete at best. For example, a decade prior to writing this review
the most popular theory to explain the population of bulges was major
mergers. At present, this is no longer an ubiquitously accepted
theory, rather it is thought by many that some mixture of turbulent
clump instabilites early on and secular evolution in more recent
epochs combine to generate many bulge propoerties
\citep[e.g.~][]{elmegreen2008,genzel2008,obreja2013}.

Below, we summarize the empircally-determined properties of
pseudobulges and subsequentially classical bulges. A very important
feature is that pseudobulge properties are not always the complement
of classical bulge properties. For example, if a bulge is star forming
(and there is no interaction present) this is very good evidence that
the bulge is a pseudobulge, but when the bulge is not star forming
this does not imply the bulge is classical. It could be either
pseudobulge or classical bulge.

The diagnostics are divided into two categories. Those in the top
catergories (or category I diagnostics) are properties in which the
parameter shows a relatively clean seperation between almost all
pseudobulges and classical bulges. The category II diagnostics are
those in which a range in parameter is only occupied by a single bulge
type, however does not identify the whole population of bulges. If one
wishes to statistically identify all bulges of a certain type in a
sample, then a category I diagnostic should be used. Alternatively, if
one has a single galaxy, or a sample of bulges and simply wishes to
know if these are bulges of a certain type then category II
diagnostics may be sufficient. For classical bulges there is a third
category which are necessary, but not sufficient properties of
classical bulges.

\subsection{Observational Definition of Pseudobulges}

Here we list the empirically-determined properties associated with
bulges that resemble disks, i.e.~pseudobulges.  \\

\noindent I - {\bf Optical morphology} in the region where the bulge
light is dominant shows spiral or ring stucture, when measured at high
spatial resolution (FWHM$\leq 100$~pc). A description of this can be
found
in \SS 2. \\
I - {\bf S\'ersic index} of bulge stellar light profile in a
bulge-disk decomposition is less than 2. Both
\cite{fisherdrory2008,fisherdrory2010}, also Fig.~\ref{fig:sersic} of
this review, show that the turnover in the distribution between
classical bulges and pseudobulges is at $n_b\sim 2.1$, and below
$n_b=2$ almost no classical bulges are observed. \\
I - {\bf Correlations with absorption line strengths} are very well
connected to bulge types \citep{peletier2007,ganda2007}. As we show in
Fig.~\ref{fig:mg}, a bulge is a pseudobulge if $\Delta$Mg~b $<$ 0.7
\AA compared to either the correlations of Mg-$\sigma$ or Mg-Fe. Below
we discuss how the
absolute value of absorption correlates with bulge type. \\
I- {\bf Velocity dispersion profile shape} thus far is the best
kinematic method to identify pseudobulges and classical bulges
\citep{fabricius2012}.  A bulge is identified as a pseudobulge if the
logarithmic derivative of the velocity dispersion profile is greater
than $dlog(\sigma)/dlog(r) \geq -0.1$ and $<v^2>/<\sigma^2> \geq
0.35$. An extreme version of this result are the so-called
$\sigma$-drop galaxies which have a local minumum in velocity
dispersion that is located where the bulge is, these galaxies would
have a positive value for $dlog(\sigma)/dlog(r)$, and thus be
pseudobulges.
\\   \centerline{ \rule{0.66\textwidth}{0.4pt} } \\
II - {\bf Low surface brightness outliers from scaling relations} are
found to be pseudobulges
\citep{carollo2001,gadotti2009,fisherdrory2010}. However, many bulges
that are co-located with fundamental plane projections also show evidence
of being pseudobulges (low $n_b$, high $SFR/M_{star}$,
\citealp{fisherdrory2010} and Fig.~\ref{fig:re_mue_sf_test}. If a
bulge is co-located with a projection of the fundamental plane, then
this does not discriminate
between being a pseudobulge or a classical bulge.\\
II - {\bf Specific star formation rate} can be indicative of
bulge types. If the region in which the bulge dominates the light has
$SFR/M_{star} \geq 10^{-11}$~yr$^{-1}$ then the bulge is very likely to
be a pseudobulge \citep{fisher2006,fdf2009}. However, if the bulge is
less active, where $SFR/M< 10^{-11}$~yr$^{-1}$, the bulge could be
either a pseudobulge or classical bulge. Care should be take also to
determine if the galaxy is presently experiencing an interaction, in
such cases correlations between $SFR/M_{star}$ and other parameters,
such as
$n_b$ become less robust. \\
II - {\bf Absorption line strength} a bulge is found to be a
pseudobulge if Fe5150 $<$ 3.95 \AA and/or Mg~b $<$ 2.35 \AA. In the
sample of SAURON based observations presented in Fig.~\ref{fig:mg}
\citep{peletier2007,ganda2007}, no classical bulge is found with
absorption lines below this range. However, this selection does not
include all pseudobulges, and therefore in a statistical study should
be used in combination with
other diagnostics. \\
II - {\bf Low - $\sigma$ outliers} to the \citet{faber1976} relation
between bulge magnitude and central velocity dispersion of the bulge
are found by \citet{kk04} to be pseudobulges. However, if a bulge is
co-located with the \citet{faber1976}, we cannot determine - from
this information alone - if it is a pseudobulge or classical bulge. \\
II - {\bf Extremely blue optical colors} statistically speaking
optical color does not appear to be a good indicator of bulge type,
however the small subset of bulges with very blue optical colors $B-V
< 0.5$ are found to be pseudobulges, and classical bulges are rare for
$B-V < 0.65$.

\subsection{Observational Definition of Classical Bulges}

We note again that classical bulges are not always the complement of
pseudobulges. In some parameter spaces there is significant overlap
between the two populations. This could be evidence of a bridging
population, but its also very likely that not every metric of galaxy
properties is uniquely manifested by a single galaxy evolution
mechanism.

The obvious condition is that first a classical bulge must not satisfy
any of the criteria listed under the definition of pseudobulges. 

\noindent I- {\bf Optical Morphology} is found to be simple and free of spiral
arms and nuclear rings in the region of the galaxy where the bulge
dominates the light. It is important to have good resolution,
preferably in the middle of the optical wavelength range ($\sim V$
through $I$ bands). In all but the closest galaxies HST is necessary
to diagnose bulge with their morphology. 
\\
I- {\bf S\'ersic Index} of classical bulges is found to be almost
always greater than two, $n_b(classical) > 2$
\citep{fisherdrory2008}. \\
I - {\bf Correlations between absorption line strengths} that are
consistent with E galaxies is a property exclusively of classical
bulges. Pseudobulges establish correlations that are offset toward
lower equivalent widths of absorption. \\
I - {\bf Strongly centrally peaking velocity dispersion profiles} are a
property that appears to be exclusively that of classical
bulges. \citet{fabricius2012} finds that if a bulge has a logaritmic
derivative that is more negative than $d log(\sigma)/dlog(r) < -0.1$
the bulge is a classical bulge. 
\\   \centerline{ \rule{0.66\textwidth}{0.4pt} } \\
II- {\bf Central Velocity Dispersion} of pseudobulges is
systematically lower than that of classical bulges. If a bulge is
found to have $\sigma_0>130$~km~s$^{-1}$ then that bulge is very
likely to also show evidence of being a classical bulge, and is not
likely a pseudobulge. However, a significant number of classical
bulges have lower $\sigma_0$ than this.

The following criteria must be satisfied to be a defined, empirically
as a classical bulge, but are not sufficient on their own to identify
the bulge as a classical bulge.
\\
III - Classical bulges are {\bf Consistent with the Fundamental plane
  scaling relationships.}\\
III - {\bf Low specific star formation rates} and {\bf low central gas
  surface densities} are found in all classical bulges, that are not
presently expereincing a merger. To be identified as a classical bulge
we find that $SFR/M_{\star}< 10^{-11}$~yr$^{-1}$ and $\Sigma_{mol} <
100$~M$_{\odot}$~pc$^{-1}$. Though many pseudobulge also have low star
formation activity and likewise are gas poor, therefore an inactive
ISM is not sufficient to identify a bulge as being either classical or
pseudobulge. \\
III - {\bf Classical bulges are not extremely blue.} There is no range
in optical color that uniquely isolates classical bulges, however if a
bulge is extremely blue it is not likely a classical bulge. \\ \\

{\bf Acknowledgments} 
DBF acknolwedges support from Australian Research Council (ARC)
Discovery Program (DP) grant DP130101460. We are grateful to
D.~Gadotti, and P.~Erwin for making data and results available to us.

\bibliographystyle{apj}

\end{document}